\documentclass[twocolumn,superscriptaddress,nofootinbib,floatfix,aps,prb,citeautoscript,longbibliography,10pt]{revtex4-2}
\usepackage[utf8]{inputenc}
\usepackage[T1]{fontenc}
\usepackage{lineno}

\usepackage{graphicx}
\usepackage{color}
\usepackage{array}
\usepackage{dcolumn}
\usepackage{bm}
\usepackage{multirow}
\usepackage{chemformula}

\usepackage{booktabs}
\usepackage{tabularx}
\usepackage{comment}
\usepackage{tikz}
\usepackage{xstring}

\usepackage{graphicx}
\usepackage{amsmath}
\usepackage{float}
\usepackage{orcidlink}
\usepackage[capitalise]{cleveref}

\begin{document}

\newcommand{\bochum}{Research Center Future Energy Materials and Systems of the University Alliance Ruhr and Interdisciplinary Centre for Advanced Materials Simulation, Ruhr University Bochum, Universitätsstraße 150, D-44801 Bochum, Germany}
\newcommand{\SCM}{Software for Chemistry \& Materials BV, De Boelelaan 1109, 1081HV Amsterdam, The Netherlands}
\newcommand{\VU}{Vrije Universiteit Amsterdam, De Boelelaan 1105, 1081 HV Amsterdam, The Netherlands}

\newcommand{\antoine}[1]{\textcolor{orange}{#1}}

\title{Universal Machine Learning Potential for Systems with Reduced Dimensionality}

\author{Giulio Benedini\,\orcidlink{0009-0000-0943-8744}}
\affiliation{\SCM}
\affiliation{\VU}
\author{Antoine Loew\,\orcidlink{0009-0008-5018-4895}}
\affiliation{\bochum}
\author{Matti Hellström\,\orcidlink{0000-0003-3053-5658}}
\affiliation{\SCM}
\author{Silvana Botti\,\orcidlink{0000-0002-4920-2370}}
\email{Corresponding author: silvana.botti@rub.de}
\author{Miguel A. L. Marques\,\orcidlink{0000-0003-0170-8222}}
\affiliation{\bochum} 

\date{\today}

\begin{abstract}
We present a benchmark designed to evaluate the predictive capabilities of universal machine learning interatomic potentials across systems of varying dimensionality. Specifically, our benchmark tests zero- (molecules, atomic clusters, etc.), one- (nanowires, nanoribbons, nanotubes, etc.), two- (atomic layers and slabs) and three-dimensional (bulk materials) compounds. The benchmark reveals that while all tested models demonstrate excellent performance for three-dimensional systems, accuracy degrades progressively for lower-dimensional structures. The best performing models for geometry optimization are orbital version 2, equiformerV2, and the equivariant Smooth Energy Network, with the equivariant Smooth Energy Network also providing the most accurate energies. Our results indicate that the best models yield, on average, errors in the atomic positions in the range of 0.01--0.02~\AA\ and errors in the energy below 10~meV/atom across all dimensionalities. These results demonstrate that state-of-the-art universal machine learning interatomic potentials have reached sufficient accuracy to serve as direct replacements for density functional theory calculations, at a small fraction of the computational cost, in simulations spanning the full range from isolated atoms to bulk solids. More significantly, the best performing models already enable efficient simulations of complex systems containing subsystems of mixed dimensionality, opening new possibilities for modeling realistic materials and interfaces.
\end{abstract}
\maketitle

\section{Introduction}
The accurate modeling of interatomic interactions remains a central challenge in computational materials science and chemistry. Traditional approaches have often faced a fundamental dilemma: quantum mechanical methods offer high accuracy but at prohibitive computational costs, while classical force fields provide efficiency at the expense of accuracy and generalizability. However, this dilemma is currently being challenged by recently developed methods such as machine learning interatomic potentials (MLIPs)~\cite{behler_perspective_2016, schmidt_recent_2019}, delivering ab-initio accuracy at computational costs comparable to classical force fields~\cite{Wang2024}. The promise of such MLIPs lies in their potential applicability to diverse problems, such as large-scale molecular dynamics with high precision, or high-throughput approaches for materials discovery and characterization.

Recently, \textit{universal} MLIPs (uMLIPs) have gained significant attention for their ability to model diverse chemical systems without requiring system-specific training~\cite{chenUniversalGraphDeep2022}. In the past couple of years, several successful uMLIPs demonstrated their capabilities in predicting energies and forces across a wide range of molecular and materials systems~\cite{Park2024, batatia_mace_2023_arxiv, chenUniversalGraphDeep2022, dengCHGNetPretrainedUniversal2023, Barroso2024, Neumann2024, mattersim}. The high transferability of these potentials originates from training on extensive datasets encompassing the whole periodic table and multiple structural motifs, enabling these models to capture complex quantum-mechanical effects. 

To assess the performance and limitations of uMLIPs, several benchmark datasets and evaluation frameworks have been developed~\cite{ Riebesell_matbench, fungBenchmarkingGraphNeural2021,Focassio2024,Haochen2024}. However, existing benchmarks tend to evaluate specific properties and systems in isolation, sometimes overlooking the importance of assessing the universal capabilities of the models. Here we turn our attention to an important element of universality, specifically how uMLIPs behave going from bulk compounds to molecules and atomic clusters, including nanowires and two-dimensional atomic layers. The transferability between spatial dimensions in essential for the study of physical systems that combine different components of different dimensionalities. A few examples are catalytic reactions at metallic surfaces, surface wetting, dissolution, combustion, crystal growth, etc. In each of these cases, a consistent and accurate description of each of components, as well of their interaction, is fundamental.

We note that the training sets of these uMLIPs frequently exhibit significant biases toward specific structural dimensions and types. For instance, large material database such as the Materials Project (MP) ~\cite{jainCommentaryMaterialsProject2013} or Alexandria (Alex) ~\cite{schmidtImprovingMachinelearningModels2024} are strongly bias toward three-dimensional (3D) crystalline structures. Similarly, molecular datasets such as ANI-2x~\cite{devereux2020}, SPICE-v2~\cite{Eastman2023,eastmanNutmegSPICEModels2024}, and QCML~\cite{Ganscha2025} contain a very specific subset of molecules (zero-dimensional, 0D) systems and are aimed to be used for specific applications. For example, the ANI-2x dataset contains seven different chemical elements, resulting in a low coverage of the chemical space in 0D.

Special attention must also be paid to the consistency of ab initio calculations across different datasets used for training and evaluating uMLIPs. It is common that various datasets are computed using different exchange-correlation functionals and computational parameters, potentially introducing systematic discrepancies in the dataset. This inconsistency becomes particularly pronounced when comparing molecular systems typically calculated with hybrid functionals such as B3LYP~\cite{Becke1993,Lee1988} against predictions from uMLIPs trained on PBE~\cite{Perdew1996} data. The energetic differences between these functionals can be substantial, leading to misleading evaluation metrics and compromising the transferability assessment of the models. 

In this work, we present a comprehensive benchmark of multiple uMLIPs across all dimensionalities from 0D (molecules, atomic clusters, etc.), passing by 1D (nanowires, nanoribbons, nanotubes, etc.) and 2D (atomic layers and slabs) to 3D (bulk materials). The multi-dimensional test systems developed for this study maintain consistent computational parameters with one of the largest training datasets employed in uMLIP development~\cite{schmidtImprovingMachinelearningModels2024}, ensuring consistency in the benchmark. Our results reveal that most modern uMLIPs exhibit a systematic reduction in predictive accuracy as dimensionality decreases, though others maintain a relatively consistent performance across all dimensional regimes.

\section{Results and Discussion}
\label{sec:results}

\subsection{uMLIPs}
\label{sec:train}

\begin{table*}[htb!]
 \caption{The uMLIP selected for this benchmark study, ordered alphabetically. We also show the targets used during training (EFS$_\textrm{G}$ or EFS$_\textrm{D}$), where E is the energy, F are the forces, S is the stress, and D and G denote if the gradients are predicted directly (D) leading to a non-conservative model or using the analytic gradient (G) resulting in a conservative uMLIP; the number of frames in the training set ($N_\text{training}$); the datasets used for the training; and the tag we use to denote the model. The data is taken from Matbench Discovery leader board~\cite{Riebesell_matbench} and from the references in the last column.} 
\label{tab:umlip_info}

\begin{tabular}{lllllll}
\toprule
uMLIP name & $N_w$ & Targets & $N_\text{training}$ & Training Datasets & Tag & Ref \\
\midrule
DPA3-v1-openlam & 8.2M & EFS$_\text{G}$ & 163M & sAlex,MPtrj,OpenLAM & DPA3 & ~\cite{zengDeePMDkitV3MultipleBackend2025} \\
eqV2-m-omat-salex-mp & 87M & EFS$_\text{D}$ & 102M & MPtrj,OMat24 & eqV2 & ~\cite{Barroso2024} \\
eSEN-30m-oam & 30M & EFS$_\text{G}$ & 113M & sAlex,MPtrj,OMat24 & eSEN & ~\cite{esen} \\
GRACE-2l-oam & 13M & EFS$_\text{G}$ & 113M & sAlex,MPtrj,OMat24 & GRACE & ~\cite{Bochkarev2024} \\
M3GNet & 0.23M & EFS$_\text{G}$ & 0.19M & MPF & M3GNet & ~\cite{chenUniversalGraphDeep2022} \\
MACE-mpa-0 & 9.1M & EFS$_\text{G}$ & 12M & sAlex,MPtrj & MACE & ~\cite{batatiaFoundationModelAtomistic2024a} \\
MatterSim-v1-5m & 4.5M & EFS$_\text{G}$ & 17M & MatterSim & MatterSim & ~\cite{mattersim} \\
ORB-v2 & 25M & EFS$_\text{D}$ & 32M & Alex,MPtrj,DDM & ORB-2 & ~\cite{Neumann2024} \\
ORB-v3-conservative-inf-mpa & 26M & EFS$_\text{G}$ & 133M & Alex,MPtrj,OMat24,DDM & ORB-3c & ~\cite{rhodesOrbv3AtomisticSimulation2025} \\
ORB-v3-direct-inf-mpa & 26M & EFS$_\text{D}$ & 133M & Alex,MPtrj,OMat24,DDM & ORB-3d & ~\cite{rhodesOrbv3AtomisticSimulation2025} \\
SevenNet-mf-ompa & 26M & EFS$_\text{G}$ & 113M & sAlex,MPtrj,OMat24 & SevenNet & ~\cite{kim_sevennet_mf_2024} \\

\bottomrule
\end{tabular}

\end{table*}

We selected 11 uMLIPs models as reported in Table~\ref{tab:umlip_info}. The names of the uMLIPs try to follow the Matbench Discovery nomenclature~\cite{Riebesell_matbench}. Most of the models are characterized by a number of parameters in the order of 20--30~million and a number of training data points in the order of several hundred million. M3GNet~\cite{chenUniversalGraphDeep2022} is included as it represents one of the first attempts at developing uMLIPS, resulting in a model with a relatively small number of parameters and training structures compared to later developments. Among the orbital (ORB) family of universal potentials~\cite{Neumann2024, rhodesOrbv3AtomisticSimulation2025} we selected ORB-v2~\cite{Neumann2024}, and their recently released ORB-v3-direct-inf and ORB-v3-conservative-inf~\cite{rhodesOrbv3AtomisticSimulation2025}. ORB-v2 is built on top of the Graph Network Simulator~\cite{sanchez-gonzalezLearningSimulateComplex} with further modifications on the architecture to leverage smoothness of the messages updates. This architecture is also characterized by the direct prediction of the forces and stresses, yielding a non-conservative model. The ORB-v3 models are designed to improve inference speed and are trained on the larger, more diverse OMat24 dataset~\cite{Barroso2024}. We chose a conservative (ORB-v3-conservative) and non-conservative (ORB-v3-direct) model with no restriction on the number of neighbors. The uMLIP eqV2-m-omat-salex-mp, another non-conservative model, is characterized by an equivariant transformers model with architectural improvements to reduce the computational costs associated to the equivariant architecture itself~\cite{Barroso2024}. The remaining models selected are all conservative. eSEN~\cite{esen} (equivariant Smooth Energy Network) takes inspiration from the EquiformerV2 architecture with a focus on smooth node representations. SevenNet~\cite{Park2024,kim_sevennet_mf_2024} extends the Nequip~\cite{nequip} framework for scalable simulations. GRACE~\cite{Bochkarev2024} is built on top of ACE descriptors~\cite{Drautz2019}, similarly to MACE~\cite{batatia_mace_2023_arxiv}, and uses an equivariant message passing architecture. The MatterSim uMLIP~\cite{mattersim} is an invariant graph neural network which is strongly influenced by M3GNet architecture, and is the second lowest in terms of number of parameters and training data. Finally, the DPA3-v1-OpenLAM model belongs to the Deep Potential with Attention (DPA) model series. This framework has evolved through successive iterations, with DPA-1~\cite{dpa-1} establishing the foundational architecture, DPA-2~\cite{dpa-2} incorporating multi-task learning capabilities to enhance transferability across diverse downstream tasks, and finally leading to the recent DPA3-v1-OpenLAM~\cite{zengDeePMDkitV3MultipleBackend2025}.

\subsection{Training datasets}

Several datasets have been used for the training of the models, are reported in Table~\ref{tab:umlip_info}. The MPF dataset~\cite{chenUniversalGraphDeep2022} used for M3GNet consisted of around 188k structures sampled from the relaxation trajectories in the Materials Project database~\cite{jainCommentaryMaterialsProject2013}. This was then expanded by including further cleaned relaxation trajectories of the Materials Project, leading to the MPtrj dataset consisting of 1.5M structures~\cite{dengCHGNetPretrainedUniversal2023}. Another popular dataset is Alex, it is derived from relaxation trajectories present in the Alexandria database ~\cite{schmidtImprovingMachinelearningModels2024}, and includes more than 30.5M data points. There exists also a sub-sampled version of the Alex dataset (sAlex)~\cite{Barroso2024}, containing approximately 10 million structures, constructed to remove the overlap with the Wang-Botti-Marques (WBM) test set~\cite{wangPredictingStableCrystalline2021} used in Matbench Discovery~\cite{Riebesell_matbench}, and to decrease the oversampling in certain regions of materials space. The OMat24 dataset~\cite{Barroso2024} extends Alexandria with more out-of-equilibrium regions of materials space, and includes 118M structures obtained through molecular dynamics runs or structural deformations. We should note that many uMLIPs adopted a two step training strategy, first by training on off-equilibrium structures (OMat24) and then by fine-tuning on close to equilibrium structures (MPtrj, Alex or sAlex). This also favours compatibility and consistency for benchmark purpose on the WBM test set. For the ORB models, there is also a zero phase where they are trained as denoising diffusion model on a dataset of relaxed structures (referred as DDM in Table~\ref{tab:umlip_info})~\cite{Neumann2024}. The DPA models (DPA-2, DPA-3) are pre-trained in the OpenLAM dataset which integrates multiple datasets with a total of more than 162 million entries (containing OMat24, MPTraj, Alex2D, SPICE2~\cite{eastmanNutmegSPICEModels2024} and many more \href{https://aissquare.com/datasets/detail?pageType=datasets&name=LAMBench-TrainingSet-v1&id=308}{[OpenLAM-v1 link]}). Finally, the MatterSim training set~\cite{mattersim} is a large-scale materials simulation dataset that includes MPTrj, Alex, and structures generated using MatterGen~\cite{zeniMatterGenGenerativeModel2024}), and that was extended with off-equilibrium ones via molecular dynamics across a wide range of temperatures and pressures.

\subsection{The 0123D dataset}
\label{sec:dataset}

\begin{figure}[htb!]
    \centering
    \includegraphics[width = 0.5 \textwidth]{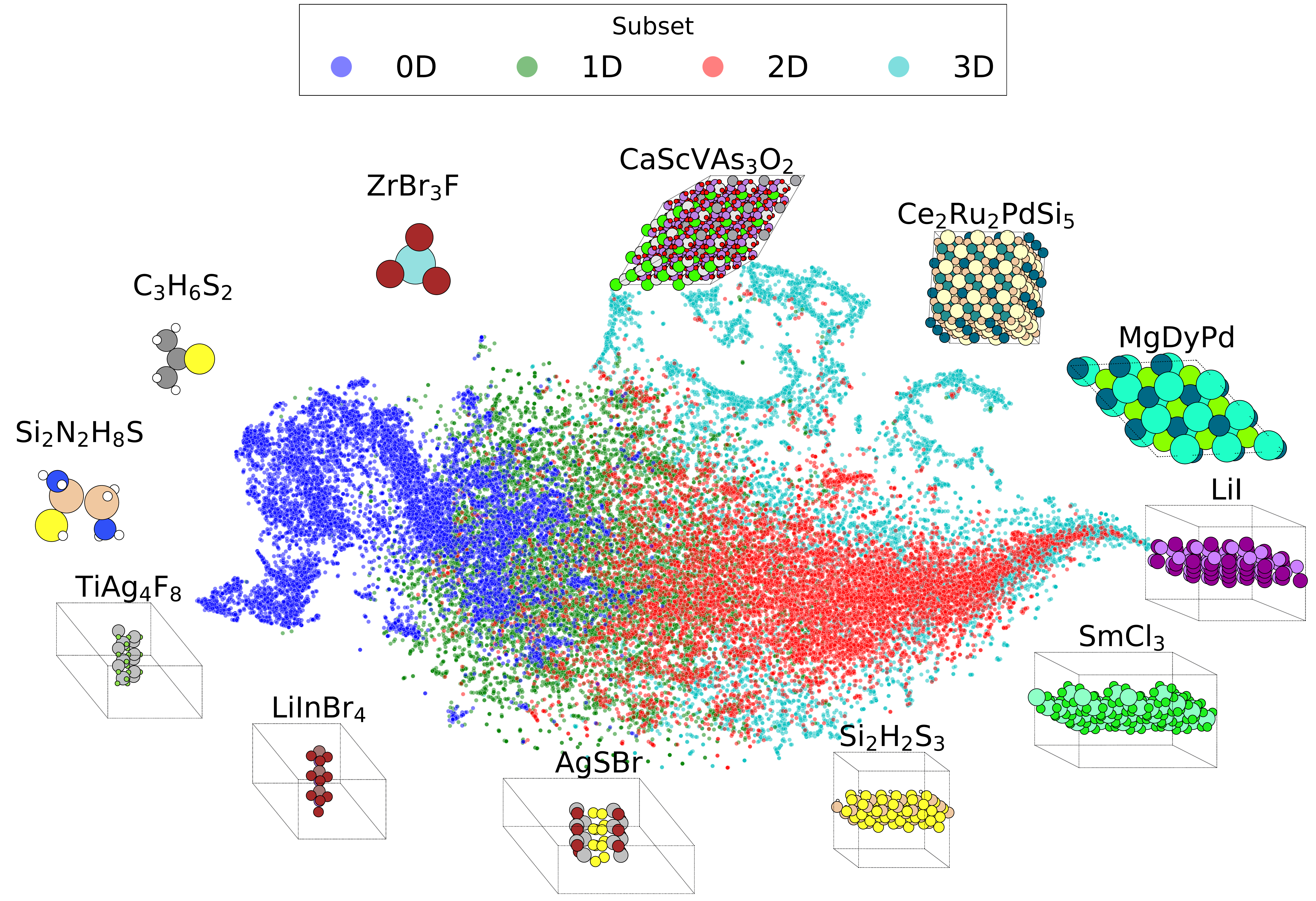}
    \caption{Scatter plot projection of the 0123D dataset using t-distributed stochastic neighbor embedding dimensionality reduction on the atomic distances distribution feature vector. Displayed are also three examples for each subsets: clockwise from the top: 3D, 2D, 1D, 0D structures. To emphasize the periodicity the atoms are repeated 3 times along the periodic directions.}
    \label{fig:fig_TSNE}
\end{figure}

\begin{figure}[htb!]
    \centering
    \includegraphics[width = 0.5 \textwidth]{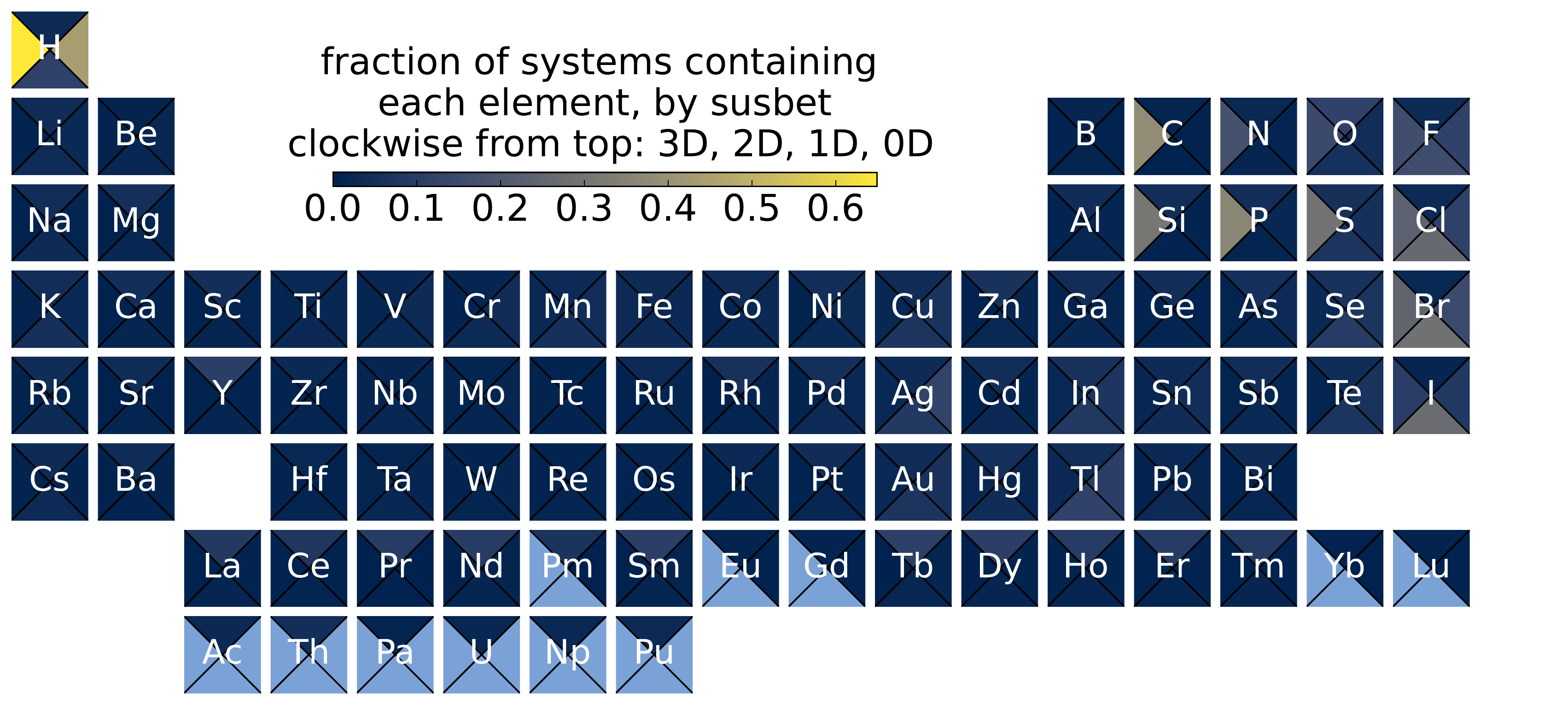}
    \caption{Fraction of systems containing a specific element of the periodic table in the 0123D dataset by dimensionality. The missing elements per subset are in light blue color. Created with pymatviz~\cite{riebesell_pymatviz_2022}.}
    \label{fig:element_presence}
\end{figure}

In this paper we introduce the 0123D dataset. Figure \ref{fig:fig_TSNE} displays the full dataset as a t-distributed stochastic neighbor embedding of the atom-distance distributions for every atomistic system, accompanied by representative structural examples. The detailed methodology for the dataset construction is described in~\cref{sec:method}. The dataset consists of 10\,000 relaxed compounds for each dimensionality, for a total of 40\,000 systems with optimized geometry and energy at the Perdew-Burke-Ernzerhof (PBE)~\cite{Perdew1996} level of theory. These compounds were chosen to be close to thermodynamic stability whenever possible, and to avoid overlap with existing training sets and the WBM dataset. These constraints had several implications in the chemical and structural variety of the 0123D dataset as we will see below.

The chemical composition of the dataset is reported in Fig.~\ref{fig:element_presence}. The 0D, 1D and 2D subsets contains elements from a large portion of the periodic table up to Bi, while for the 3D subset this was extended further to include Po, At, and the actinides up to Pu. The distribution of the elements shows some deviations from a uniform sampling: the 0D subset presents approximately 3000 systems that contain H, C, Si, P, S and Cl, due to inclusion of organic molecules. The 1D and 2D subsets show an excess of H, F, Cl, Br, and I, with more than 1000 systems containing one of these elements, due to the requirement of charge neutrality in the construction of the dataset (see~\cref{sec:method}). The 3D subset has a much more uniform distribution of the chemical elements, but with a stronger emphasis on the lanthanides. This over-representation of lanthanides in stable compounds is not only observed here, but also in the GNoME convex hull~\cite{merchant2023scaling}, and is ultimately caused by the strong chemical similarity between these chemical elements. Although chemical element distributions differ across dimensionalities, our dataset is sufficiently large and representative to yield robust and meaningful conclusions.

\begin{figure}[htb!]
    \centering
    \includegraphics[width = 0.47 \textwidth]{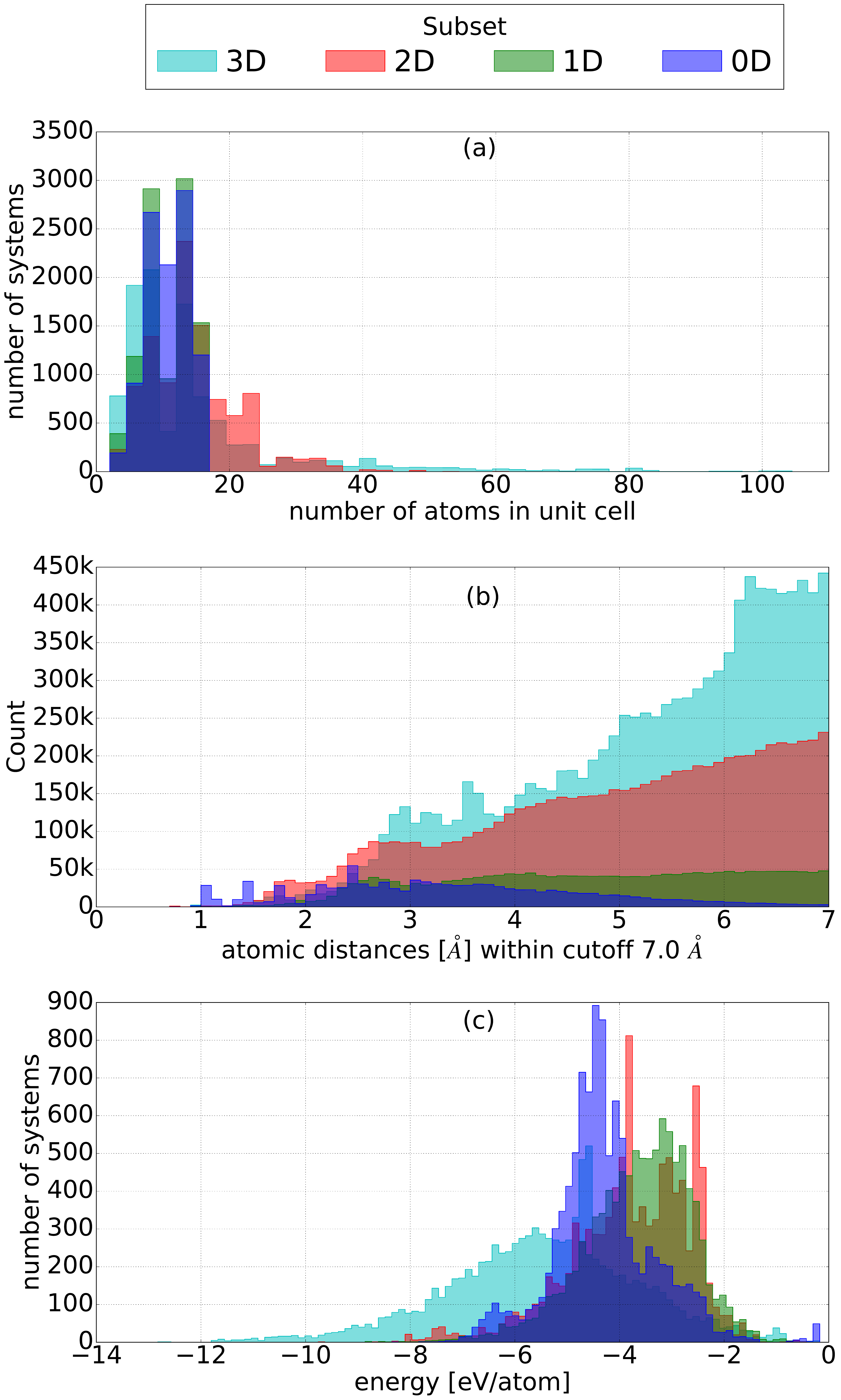}
    \caption{Distributions of (a)~number of atoms, (b)~the distances between pairs of atoms within a 7~\AA\ cutoff radius and (3)~the total energy per atom for 0123D dataset as a function of dimensionality.}
    \label{fig:property_distribution}
\end{figure}

In \cref{fig:property_distribution} we plot the distribution of the number of atoms, atomic distances, and energy per atom as a function of dimensionality. In panel (a) we can see that most compounds contain less than 20 atoms in the primitive unit cell, although for 2D and 3D this number can be higher, reaching 100 atoms per unit cell for the 3D case. The smaller number of atoms for low dimensionality is related to the increased computational costs due to the inclusion of vacuum in the unit cell.

Panel (b) has to be read carefully due to the strong dependence of the atomic distances with dimensionality. In fact, for a given atom, the number of neighbors in a N-dimensional shell of inner radius $R$ goes to zero for 0D when $R$ is larger than the diameter of the system, goes to a constant for 1D, as $R$ for 2D, and as $R^{2}$ for 3D. Therefore, the trends in the distribution, as shown in the panel (b) goes like the derivative of the before mentioned trends. The sharp peaks in the 0D curve are due to the short covalent bonds between the first-row atoms that compose the organic molecules. On the other hand, the large peak starting at around 2.5~\AA\ comes from the longer bonds in compounds with chemical elements from later periods.

Finally, in the bottom panel of \cref{fig:property_distribution} we plot the energy per atom of the different compounds. We emphasize that the total energy does not have a physical meaning, but it is well defined within a given numerical approach, and is important to benchmark uMLIPs. Most compounds have energies per atom between -2 and -6~eV/atom, in particular for the 012D case. The 3D compounds have on average lower energies, which is not surprising as they do not possess surface atoms that are typically under-coordinated and that therefore lead to dangling bonds. The 0D systems also appear on average at lower energies than 1D or 2D. 

\subsection{Benchmark}
\label{sec:test}

\begin{table}
 \caption{Number of systems that failed to converge during the geometry optimization, for each uMLIP as a function of dimensionality. The uMLIPs are listed in alphabetic order.} 
 \label{tab:not_converged}
\begin{tabular}{lrrrr}
\toprule
uMLIP & 0D & 1D & 2D & 3D \\
\midrule
DPA3 & 25 & 2 & 49 & 8 \\
eqV2 & 620 & 80 & 35 & 3 \\
eSEN & 0 & 0 & 32 & 3 \\
GRACE & 0 & 0 & 56 & 5 \\
M3GNet & 1 & 0 & 33 & 5 \\
MACE & 0 & 0 & 47 & 6 \\
MatterSim & 0 & 0 & 42 & 5 \\
ORB-2 & 1 & 2 & 16 & 2 \\
ORB-3c & 0 & 0 & 34 & 8 \\
ORB-3d & 89 & 26 & 58 & 10 \\
SevenNet & 0 & 0 & 43 & 6 \\
\bottomrule
\end{tabular}
\end{table}

There are several possible metrics to measure the performance of uMLIPs with respect to the reference data. The simplest are perhaps the number of failed relaxations and the number of relaxation steps required for convergence to the minimum energy structure (with respect to our converge threshold). The number of systems that failed to converge for each dimensionality subset is reported in Table~\ref{tab:not_converged}. It turns out that most uMLIPs manage to achieve convergence for the overwhelming majority of the structures. However, we can detect two notorious exceptions, specifically eqV2 and ORB-3d, the two non-conservative uMLIPs in our study, where the number of unconverged relaxations is very high. This behavior is likely due to small high-frequency errors in the direct prediction of the forces that complicates considerably the geometry relaxation process. Curiously, this behavior is absent from ORB-2, meaning that the problem related to non conservative forces can be considerably alleviated.
Finally, the larger number of failures in 2D is related to some multi-layered systems that upon uMLIP relaxation exceed our thickness threshold of 7.5~\AA (see the methods section for details).

\begin{figure}[htb!]
    \centering
    \includegraphics[width = 0.47 \textwidth]{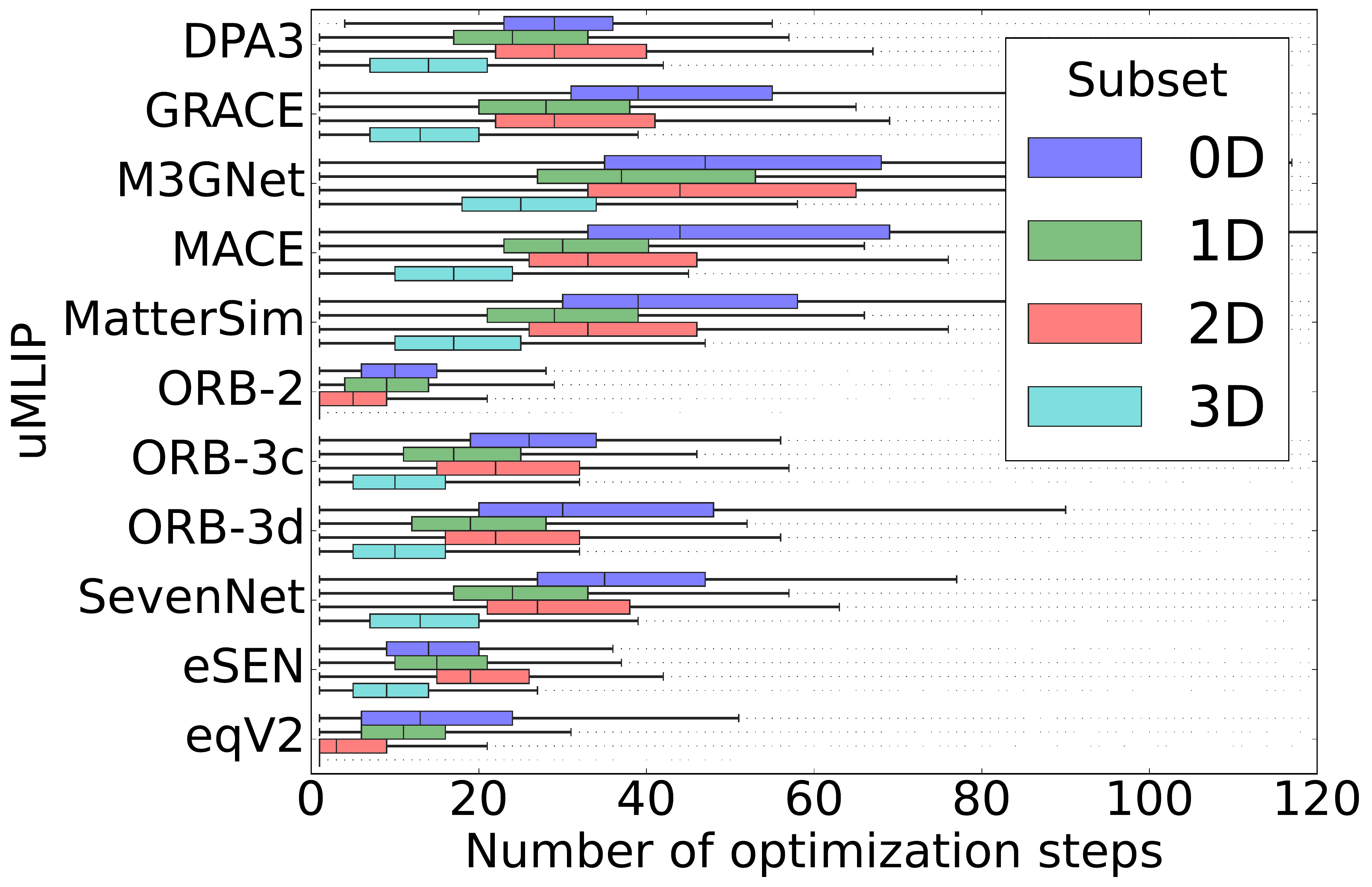}
    \caption{The box plot distribution of the number of steps to converge the geometry optimization calculation for each uMLIP per dimensionality subset. The starting structures are the one present in 0123D dataset.}
    \label{fig:optimization_setps}
\end{figure}

Additional insight can be gained from analyzing the number of optimization steps required for convergence, as shown in Fig.~\ref{fig:optimization_setps}. For 3D the number of relaxation steps is small, and is usually below 20 steps for most modern uMLIPs. Remarkably, for eqV2 and ORB-2 most compounds are already converged to the required accuracy after 1 step, which shows the performance of these two uMLIPs close to dynamical equilibrium. We can also observe the impressive improvement of uMLIPs for the past 4 years since the introduction of M3GNet. For all uMLIPs we see a considerable deterioration of the quality of the potential energy surface for lower dimensionalities. As expected, this deterioration increases roughly with decreasing dimensionality, as we go further from the bulk systems that constitute the large majority of the systems used for training these uMLIPs. Note that the larger number of optimization steps for 2D is simply related to the larger average number of atoms (see Fig.~\ref{fig:property_distribution}(a)). The best performing model in this metric is ORB-2, followed by eqV2 and eSEN.

\begin{figure}[htb!]
    \centering
    \includegraphics[width = 0.43 \textwidth]{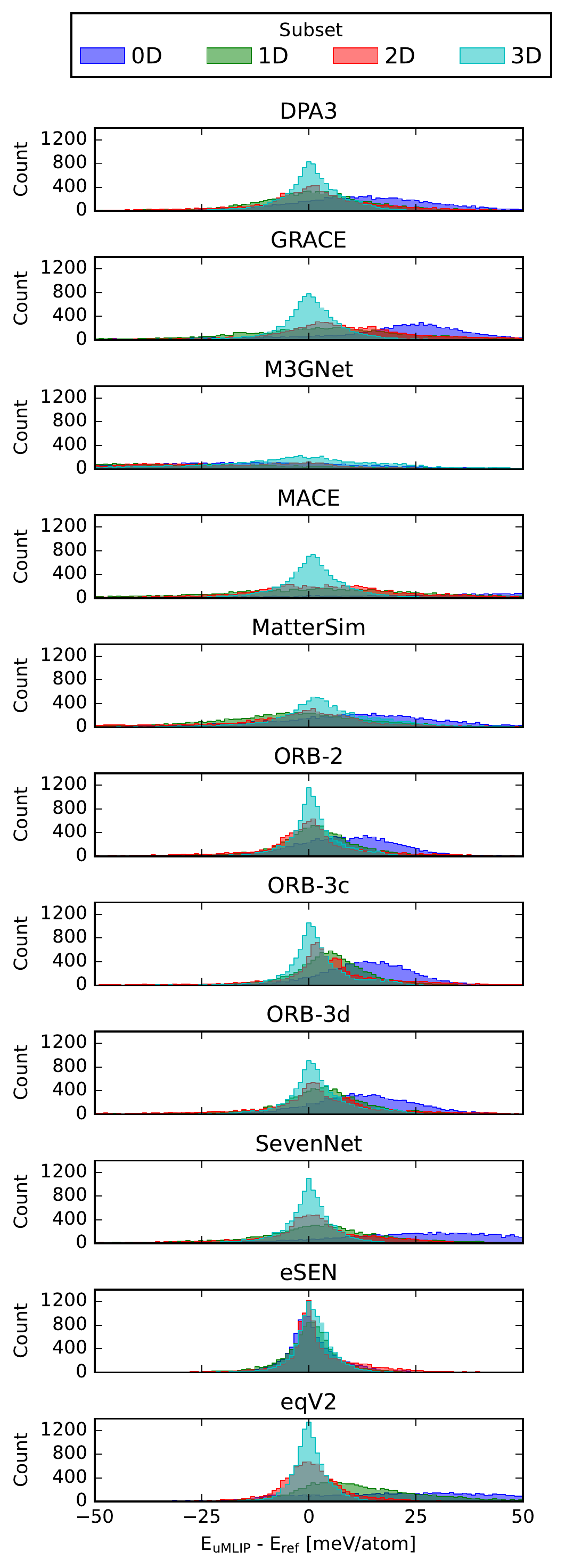}
    \caption{Distribution of energies differences per atom for each uMLIP.}
    \label{fig:energy_delta_dist}
\end{figure}

We now turn our attention to the error distribution of the energy, as shown in Fig.~\ref{fig:energy_delta_dist}. Most uMLIPs perform extremely well on the 3D subset, with the exception of M3GNet. 
The errors of the more recent uMLIPs are typically below 10 meV/atom, which is approaching the commonly referenced chemical accuracy threshold of 1 kcal/mol ($\sim$43 meV) and close to the numerical precision of the datasets (a few meV/atom).
This again confirms that modern uMLIPs are more than capable of replacing DFT codes in the simulation of bulk compounds close to dynamical equilibrium at a small fraction of the computational cost.

When moving from 3D to the lower dimensional subsets the distribution of the errors of all the models broadens. Furthermore, in most cases there is an evident systematic error, with a considerable overestimation of the energy. This behavior increases with decreasing dimensionality, as we move further away from the bulk compounds used in training.  Because most models excel on the 3D subset, they may be biased toward these structures. Three-dimensional systems place more atoms within the cutoff radius than lower-dimensional ones, which could make the latter appear less stable than they truly are. For the GRACE and the ORB models the overestimation of the energy is the range of 10--40~meV/atom, meaning that they are still useful for the study of systems of reduced dimensions. However, for DPA3, eqV2, M3GNet, MACE, MatterSim, and SevenNet the error, especially for 0D systems, is considerable, which limits the applicability of these uMLIPs. 
The DPA3 model shows unexpectedly poor performance despite its training on diverse systems that include molecular configurations. The best performing uMLIP, and therefore the most transferable, is without doubt eSEN, for which more than 75\% of the energy predictions on 0123D dataset have an error lower than 10~meV/atom.

\begin{figure}[htb!]
    \centering
    \includegraphics[width = 0.47 \textwidth]{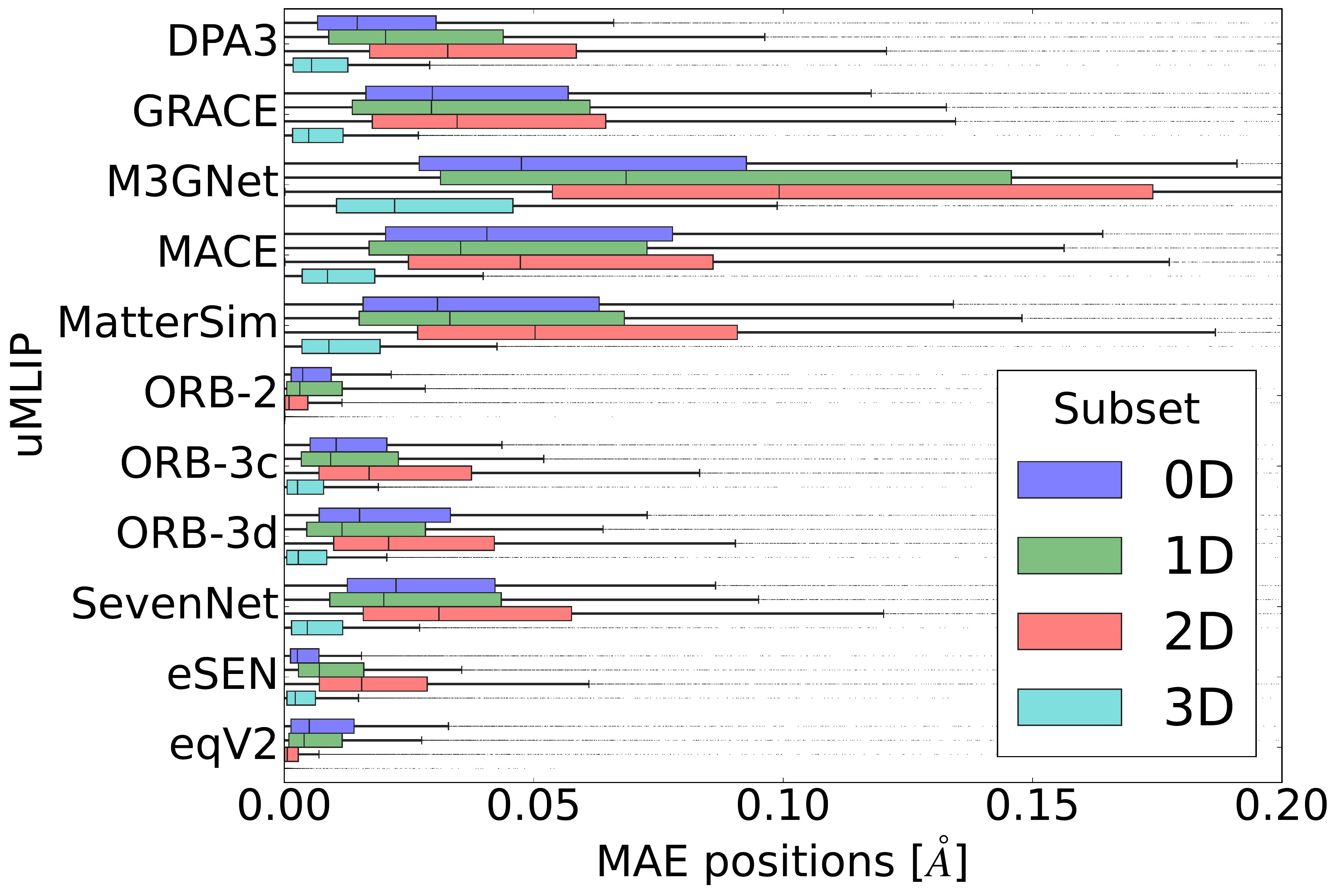}
    \caption{The box plot distribution of the mean absolute error (MAE) between atoms position for each uMLIP.}
    \label{fig:geometry_mae_box}
\end{figure}

Finally, we look at the error in the geometry in Fig.~\ref{fig:geometry_mae_box}. The trends correlates closely with the number of optimization steps required for convergence (see Fig.~\ref{fig:optimization_setps}). All models, perhaps with the exception of M3GNet, perform extremely well for 3D systems, but with a clear degradation for lower dimensions. Overall the best performing models with respect to the geometry are ORB-2 and eqV2 followed by eSEN. Interestingly, while ORB-2 and eqV2 yield the best geometries, the energy is somewhat less accurate, as can be seen in Fig.~\ref{fig:energy_delta_dist}. Curiously ORB-3 models perform consistently worse than their ORB-2 predecessor, indicating that the computational efficiency improvements implemented in the newer models came at the cost of reduced accuracy. Finally, our results demonstrate that both direct and conservative force prediction approaches can yield high quality geometries.

\subsection{Conclusion}
\label{sec:conclusion}

In conclusion, we find that most uMLIPs exhibit degraded performance when applied to systems of reduced dimensionality compared to 3D bulk compounds. This degradation stems from several factors: first, the majority of training data for these models consists of 3D systems; second, the transition from 3D to lower dimensions involves significant changes in atomic coordination and bond lengths, resulting in fundamentally different chemical and physical behavior. Therefore, some degree of performance degradation is expected when extending these models beyond their primary training domain. A notable exception is the eSEN model, which exhibits remarkable robustness with errors in atomic positions remaining consistently within 0.01–0.02~\AA\ and energy errors below 10~meV/atom across all dimensionalities. The ORB and GRACE models also demonstrate good performance in this regard. These results suggest that certain uMLIPs, particularly eSEN, are already well-suited for simulations involving subsystems of varying dimensionality and their interactions. We tentatively attribute the superior transferability of eSEN to its training strategy and pre-training methodology rather than architectural differences, as its architecture and training data are comparable to other uMLIPs. However, further investigation would be needed to definitively establish the source of this advantage. This insight provides a valuable lesson for improving the dimensional transferability of other uMLIPs.

\section{Methods}
\label{sec:method}

\subsection{Dataset}

We constructed our dataset with three key objectives in mind. First, we aimed to achieve comprehensive coverage of the periodic table, including the majority of chemical elements. Second, we sought to encompass a reasonable diversity of geometric arrangements, with particular emphasis on configurations near thermodynamic stability. Finally, we ensured that our dataset did not overlap with the systems used in the training of uMLIPs, thereby minimizing potential contamination and the associated uncertainty in our results. We recognized that the unique characteristics of different dimensionalities necessitated tailored approaches, leading us to adopt dimension-specific strategies in the construction of our dataset.

There are generally available datasets that include a wealth of DFT calculations for 3D compounds, and these are commonly used for the training of uMLIPs. To create the 3D dataset, we used the model of Ref.~\onlinecite{DeBreuck2025} to generate 3 million structures that the model believe were closed to the convex hull. These were optimized with ORB-2 model~\cite{Neumann2024} model, duplicates (compounds already present in Alexandria) were removed, and the distance to the convex hull was estimated with the ALIGNN model of Ref.~\cite{schmidtImprovingMachinelearningModels2024}. From the compounds closer to the hull we selected randomly 10\,000 entries that were further relaxed with DFT. 

For lower dimensionalities we do not have available a generative model with the same level as accuracy of the one of Ref.~\onlinecite{DeBreuck2025}. Therefore, we decided to use the PyXtal software~\cite{pyxtal} to generate compounds in random space groups and with random occupations of the Wyckoff positions. This enables a comprehensive exploration of crystallographically valid structures across different space groups. Note that to increase the probability that the generated structures are close to thermodynamic stability, this workflow imposes charge neutrality constraints. In this way we generated 2 million systems for each of the lower dimensionalities. For 0D-systems, we also decided to add the molecular structures present in the Materials Project~\cite{jainCommentaryMaterialsProject2013} database as well as computationally generated atomic clusters to ensure comprehensive coverage of isolated molecular and cluster systems.
All these initial structures were again pre-relaxed with ORB-2 model~\cite{Neumann2024}, and the distance to the convex hull was calculated using the ORB-2 energy, as we do not have at the moment a reliable model to predict directly the distance to the hull for lower dimensionality systems. The workflow then followed the same steps as for 3D, resulting in 10\,000 relaxed DFT calculations for each of the dimensionalities.

Details on the numerical procedure for the DFT calculations can be found in Ref.~\onlinecite{schmidtImprovingMachinelearningModels2024} and have been chosen to maintain consistent computational parameters with one of the largest training datasets employed in uMLIP development~\cite{schmidtImprovingMachinelearningModels2024} and the WBM test set~\cite{wangPredictingStableCrystalline2021}.

\subsection{Geometry relaxation}

Clearly the inference error is an important metric in the assessment of a uMLIP, but it does not reflect a typical workflow in materials science. Therefore we decided to calculate errors relative to the relaxed structures in the individual methods. We performed the benchmark by performing a geometry relaxation with each uMLIP starting from the optimized DFT geometry of the 0123D dataset. We used the ASE~\cite{HjorthLarsen2017} interface to the uMLIPs and the FIRE geometry optimizer~\cite{Bitzek2006}. We stopped the geometry relaxation when the forces were converged to better than 40~meV/\AA, when the number of iterations exceeded 15000, or when the force exceeded 10\,000~eV/\AA\ (indicating a serious problem in the uMLIP). In the last two situations, the structure was labeled as unconverged. For the reduced dimensions, we also imposed geometrical thresholds to detect fragmentation of the systems. Specifically we discarded 2D slabs thicker than 7.5~\AA, 1D systems wider than 12.5~\AA, and 0D systems with diameter larger than 20~\AA. 

\subsection{Comparison between geometries}

To compare atomic geometries between uMLIP predictions and reference calculations, we need a strategy to compress the comparison of two 3N atomic positions (where N is the number of atoms) and unit cell parameters into a single meaningful metric. Furthermore, we require a quantity that remains significant across all system dimensionalities, from 0D to 3D. Our geometry comparison procedure consists of two steps: First we align the structure by mapping uMLIP atomic positions to the corresponding PBE reference positions. Under the assumption that no atomic permutations occur during geometry optimization, the atom-to-atom correspondence is straightforward. To eliminate the problem of atoms wrapping across periodic boundaries, we minimize the interatomic distances in the uMLIP cell, using the PBE unit cell as the reference. We employ the Kabsch algorithm~\cite{kabsch1976solution} to align the two structures through rotation and translation. Finally, we calculate the mean absolute error between corresponding atomic position components. This metric was chosen as it is less sensitive to outliers, and it is less influenced by the total number of atoms in the system, allowing for more consistent comparisons across different system sizes.

\section{Data Availability}

The benchmark structures can be downloaded from the Alexandria database at \url{https://alexandria.icams.rub.de/}. As this is meant as a benchmark, we ask model makers not to include this data into their training or validation datasets.

\section{Code Availability}

\label{sec:code_ava}
All code used in this work is freely available at \url{https://github.com/hyllios/utils/tree/main/} and at \url{https://github.com/GiulioIlBen}.

\section{Acknowledgements}

We acknowledge funding from the Horizon Europe MSCA Doctoral network grant n.101073486, EUSpecLab, funded by the European Union. S.B. acknowledge funding from the Volkswagen Stiftung (Momentum) through the project ‘‘dandelion''. M.A.L.M would like to thank the NHR Centre PC2 for providing computing time on the Noctua supercomputers.

\bibliography{old/main.bbl}

\end{document}